\batchmode
\makeatletter
\def\input@path{{\string"C:/Documents and Settings/BTI/My Documents/References/Photons/Photon Models/Lorenz/LorenzOnLight/\string"/}}
\makeatother
\documentclass[oneside,english,reqno]{amsart}
\usepackage[T1]{fontenc}
\usepackage[latin1]{inputenc}
\usepackage{geometry}
\usepackage{setspace}
\usepackage{amssymb}

\makeatletter

\providecommand{\tabularnewline}{\\}

 \theoremstyle{plain}

\usepackage{babel}
\makeatother
\begin{document}

\title{Lorenz on Light: A Precocious Photon Paradigm}

\author{H. C. Potter}

\begin{abstract}
In 1867, during the time when Maxwell was publishing his electromagnetic
theory, L. Lorenz published his theory equating light vibrations with
electric currents. Starting from Kirchhoff's Ohm's law expression,
Lorenz introduces scalar potential retardation to obtain an expression
for the electric field using retarded potentials. In his theory Lorenz
sets the vacuum light speed equal to $\sqrt{2}/2$ times Weber's measured
superluminal value for magnetic induction speed. Using the wave equation
Green's function paradigm Lorenz reduces his integral, retarded Ohm's
law formulation to a differential formulation for current density.
This represents his solution to the electromagnetic action at a distance
problem. During the reduction he presents a light like, plane wave
solution for current density that can not satisfy initial conditions
on the expanding wave front and, using a faulty by parts integration
that neglects retardation, he develops the eponymic Lorenz condition.
Although generally accepted, this condition relates vector and scalar
potentials erroneously. A proper relation is given and major consequences
are developed in appendices. As a finale Lorenz suggests a light model
that I interpret as magnetoinductive waves. Such waves are only now
being studied using Kirchhoff's laws without retardation. I redo this
important paper using modern day notation with critical annotations.
This elevates the paper to a teaching tool position in the millennial
search to understand light's dual nature as wave and corpuscle and
brings us full circle, back to Kirchhoff's laws where Lorenz began. 
\end{abstract}

\keywords{Light theory, Lorenz condition, Maxwell equations, Photons}

\maketitle
\newpage{}

\section{Introduction}

In researching photon models I obtained the 1867 article entitled
{}``On the identity of the vibrations of light with electrical currents''
\cite{LLorenz} expecting to find the rationale for the Lorenz condition
that relates electromagnetic potentials. I found much more: wave equations
for light derived from Ohm's law by retarding scalar potential propagation
and reference to light as rotationally polarized by transversely circulating
currents. I interpret these currents as forming magnetoinductive waves
exhibiting autonomous progression. As such, these current loops represent
a precocious model for light quanta whose photon name was not proposed
\cite{Lewis_Photon} for another 59 years. The role Lorenz played
in developing electromagnetic theory and, in particular, his 1867
paper precedes gauge theory development. Gauge theory has come to
be based on the observation by H. A.Lorentz (1853-1928) that special
potential transformations leave select electromagnetic Maxwell fields
invariant \cite{Jackson&Okun}. This observation was apparently made
as an aside when Lorentz discovered that the Lorenz condition was
required for the Maxwell electromagnetism equations to be satisfied
for fields derived from wave function potentials. The Lorenz condition
inverse development from wave function potentials in \cite{LLorenz}
was found only later and led to shared attribution, but the Lorenz
development was not critically scrutinized as is done below. This
oversight resulted in a failure to detect that the Lorenz condition
development can be deemed flawed when the vector potential is treated
as a local function whose spatial derivatives at a point depend only
on function values near the point. This failure is etiologically significant;
for, if the Lorentz discovery can be inverted, Maxwell's electromagnetism
equations and, ultimately, the Lorentz transformation on which relativity
is based would follow from Ohm's law and Lorenz retardation. This
flaw is exposed below by translating the Lorenz equations to modern
vector form and explained in \ref{sub:Lorenz-Condition}. Its effect
on the field equations is presented in Appendix \ref{sec:Field-Equations}.
In this paper I give a detailed, section by section reprise with annotations
appropriate to our modern day understandings. But first some background
will help explain why this paper is important and why it was nearly
lost to posterity.

Lorenz, Ludvig Valentin, was a Danish physicist who lived from 1829-1891.
Since the article was published during the time period (1865-1873)
when Maxwell (1831-1879) was publishing his work and does not refer
to Maxwell's formulation, the mathematics needs translation into present
day terms for ready comprehension. With Maxwell's vacuum light velocity
prediction $c_{0}=1/\sqrt{\epsilon_{0}\mu_{0}}$ from the vacuum electric
and magnetic constants for transverse electromagnetic waves, his field
equations became the standard for modeling light as traveling plane
waves. These Maxwell waves carry in phase electric and magnetic fields
orthogonal to one another and the travel direction. The subsequent
discovery that light had a particle like character in the early 20$^{\text{th}}$
century posed duality reconciliation as a problem for physicists.
Attempts to resolve this problem that my search has uncovered are
succinctly summarized in Table 1 which supplements and updates the
 review in \cite{Kidd}. 

In the next section I present the Lorenz work as I interpret it using
present day vector calculus. I present the work in small, well defined
sections with most sections followed by my critique. Following this
presentation I discuss this work's educational significance more broadly.

\section{Presentation}

In this section I concisely present the Lorenz analysis using present
day symbology. The original equation numbering, sometimes unconventional,
is retained to facilitate cross checking. With some clear exceptions
many other equations represent unnumbered Lorenz relations. Since
the symbol we presently associate with full differentiation is used
throughout the work, our symbol for partial differentiation has been
introduced were required. Subsection critiques have been added to
facilitate concept binding.

\subsection{Analytical Basis}

Lorenz bases his analysis on a starting formulation that he attributes
to Kirchhoff.

\subsubsection{Kirchhoff formulation\label{sub:Kirchhoff-formulation}}

Consistent with the article title, Lorenz begins with equations that
today we would associate with Ohm's law expressed as\begin{equation}
\mathbf{J}=\kappa\mathbf{E}=-\kappa\left(\boldsymbol{\nabla}\Omega+\frac{2}{c}\frac{\partial\mathbf{A}}{\partial t}\right)\tag{L1}{}\label{eq:L1}\end{equation}
where the current density, $\mathbf{J}$, equals the conductivity,
$\kappa$, times an electric field, $\mathbf{E}$, equal to a negative
scalar field gradient, $-\mathbf{\boldsymbol{\nabla}}\Omega$, plus
a normalized negative vector potential temporal variation, $-\frac{2}{c}\frac{\partial\mathbf{A}}{\partial t}$.
Following Kirchhoff, Lorenz expressed the vector potential \cite{Jackson&Okun}
as\begin{equation}
\mathbf{A}_{K}=\frac{2}{c}\int\int\int\frac{\mathbf{R}}{R^{3}}\mathbf{R}\bullet\mathbf{J}(\mathbf{r}',t)dv'\label{eq:AsubK}\end{equation}
 and the scalar potential as Coulomb potential superposition\begin{equation}
\Omega=\int\int\int\frac{\mathbf{\mathrm{\epsilon'}}}{R}dv'+\int\frac{e'}{R}ds'\label{eq:Omega}\end{equation}
with the continuity equations for charge\begin{equation}
\left.\begin{array}{ccc}
\boldsymbol{\nabla}\bullet\mathbf{J} & = & -\frac{1}{2}\frac{\partial\epsilon}{\partial t},\\
\\\mathbf{J}\bullet d\mathbf{n} & = & -\frac{1}{2}\frac{de}{dt}\end{array}\right\} \tag{L2}{}\label{eq:L2}\end{equation}
where the $\epsilon'$ is the charge density at the relative position
$\mathbf{R}=\mathbf{r}-\mathbf{r}'$ in a Cartesian system and $e'$
is the charge density on the surface element $ds'$ with inward normal
$d\mathbf{n}$. Here the primed densities are in electric potential
units. So the electric constant with value 8.85E-12 Fm$^{-1}$, F=QV$^{-1}$,
must be introduced where required.

\subsubsection{Kirchhoff Formulation Critique\label{sub:Kirchhoff-Formulation-Critique}}

The Eqs.(\ref{eq:L2}) express charge conservation. The first follows
redundantly from the Ampere law expressed using the Maxwell magnetizing
force, $\mathbf{H}$, and electric induction, $\mathbf{D}$, as\begin{equation}
\boldsymbol{\nabla}\times\mathbf{H}=\frac{\partial\mathbf{D}}{\partial t}+\mathbf{J}\label{eq:Ampere}\end{equation}
with $\boldsymbol{\nabla}\bullet\mathbf{D}=\rho$ since $\boldsymbol{\nabla}\bullet\left(\boldsymbol{\nabla}\times\mathbf{H}\right)\equiv0$.
The second follows from $\mathbf{J=\rho\mathbf{v}}$ where $\rho$
is the local charge per unit volume and $\mathbf{v}$ is its local
velocity. When the magnetic induction is defined by\begin{equation}
\mathbf{\mathbf{B}}=\frac{2}{c}\mathbf{\boldsymbol{\nabla}\times\mathbf{A}},\label{eq:Bdef}\end{equation}
the Eq.(\ref{eq:L1}) expresses the Faraday law\begin{equation}
\boldsymbol{\nabla}\times\mathbf{E}=-\frac{\partial\mathbf{B}}{\partial t}\label{eq:Faraday}\end{equation}
 since $\boldsymbol{\nabla}\times\boldsymbol{\nabla}\Omega\equiv0$.
Thus, without explicitly invoking magnetic effects, the formulation
is implicitly consistent with Maxwell's four equations . The scalar
potential can be considered a Green's function solution for the Poisson
equation, $\nabla^{2}\Omega=-\rho$ , in a connected volume that includes
surface terms that are seldom considered in present day analyses.
The vector potential is an \emph{ad hoc} expression. Below it will
be brought surprisingly into conformity with the present day retarded
form.

The formulation used by Lorenz considers charge and current distributions
to be fundamental observables from which real potentials can be obtained.
This subverts the less than pure field Maxwell conception that real
charge and current distributions give rise to observable electromagnetic
fields but nonphysical potentials \cite[Chapter XII]{Dirac}. Contributing
to potential nonphysicality is susceptibility to transformation with
an arbitrary gauge function, $\chi$, that causes\begin{equation}
\left.\begin{array}{ccc}
\mathbf{A}\rightarrow\mathbf{A}' & = & \mathbf{A}+\boldsymbol{\nabla}\chi,\\
\Omega\rightarrow\Omega' & = & \Omega-\frac{1}{c}\frac{\partial\chi}{\partial t}.\end{array}\right\} \label{eq:GaugeTransform}\end{equation}
 Many such transformations are presented in \cite{Jackson}. Two are
given special recognition. The $\chi$ choice that gives $\boldsymbol{\nabla}\bullet\mathbf{A}=0$
is called the Coulomb gauge, while that that gives $\boldsymbol{\nabla}\bullet\mathbf{A}+\frac{1}{c}\frac{\partial\Omega}{\partial t}=0$
is called the Lorenz gauge. Expressions for the new potentials have
come to be called {}``gauges'' without regard for gauge function
existence \cite{Heras}. This custom is inappropriate, because the
gauge expressions can be satisfied by nongauge transformations. For
example, the Coulomb gauge relation can be satisfied whenever the
vector potential is augmented by some vector's curl; but, by the Helmholtz
Theorem, that augmentation can not be a gauge function's gradient.
Thus, the gauge relations actually restrict the problem relevant potentials.
This justifies the proposal in \cite{Onoochin} that electromagnetic
potential specification is essential for obtaining complete electromagnetic
field distributions. This proposal is supported by the finding in
\ref{sub:Lorenz-Condition} that there are \emph{two} distinct retarded
scalar potentials, Eq.(\ref{eq:OmegaBar}) $\overline{\Omega}$ dependent
on remote charge and Eq.(\ref{eq:OmegaBarZero}) $\overline{\Omega}_{0}$
dependent on remote current.

\subsection{Basis Extension}

Lorenz considers the Eq.(\ref{eq:L1}) right side terms to be the
first terms in series expansions that add potential retardation.

\subsubsection{Retardation\label{sub:Retardation}}

Designating the retarded time by $t-\frac{R}{a}$, Lorenz defines
the new function

\begin{equation}
\overline{\Omega}=\int\int\int\frac{\mathbf{\mathrm{\epsilon'(t-\frac{R}{a})}}}{R}dv'+\int\frac{e'(t-\frac{R}{a})}{R}ds'.\label{eq:OmegaBar}\end{equation}
The charge densities are expanded to give\begin{equation}
\left.\begin{array}{ccc}
\epsilon'(t-\frac{R}{a}) & = & \epsilon'-\frac{\partial\epsilon'}{\partial t}\frac{R}{a}+\frac{\partial^{2}\epsilon'}{\partial t^{2}}\left(\frac{R}{a}\right)^{2}\frac{1}{2}-\cdots,\\
\\e'(t-\frac{R}{a}) & = & e'-\frac{de'}{dt}\frac{R}{a}+\frac{d^{2}e'}{dt^{2}}\left(\frac{R}{a}\right)^{2}\frac{1}{2}-\cdots.\end{array}\right\} \label{eq:ChargeDensities}\end{equation}
When the expansions are inserted\begin{equation}
\boldsymbol{\nabla}\overline{\Omega}=\boldsymbol{\nabla}\Omega+\frac{1}{2a^{2}}\left[\int\int\int\frac{\mathbf{\mathbf{R}}}{R}\frac{\partial^{2}\epsilon'}{\partial t^{2}}dv'+\int\frac{\mathbf{R}}{R}\frac{d^{2}e'}{dt^{2}}ds'\right]-\cdots.\label{eq:GradOmegaBar}\end{equation}
 Careful coordinate tracking confirms development validity to this
point, but note that the integrands no longer depend on the retarded
time. Here Lorenz states that using Eqs.(\ref{eq:L2}) and partial
integration gives\begin{equation}
\boldsymbol{\nabla}\overline{\Omega}+\frac{1}{a^{2}}\frac{\partial}{\partial t}\int\int\int\frac{\mathbf{\mathbf{J'}}}{R}dv'=\boldsymbol{\nabla}\Omega+\frac{c}{2a^{2}}\frac{\partial\mathbf{A}_{K}}{\partial t},\tag{L4}{}\label{eq:L4}\end{equation}
where $\mathbf{\mathbf{J'}}$ indicates that $\mathbf{\mathbf{J}}$
depends on $t-\frac{R}{a}$ and $\mathbf{r}'$ rather than $t$ and
$\mathbf{r}$.

\subsubsection{Retardation Critique}

Despite the \ref{sub:Retardation} predominant significance for the
Lorenz paper, the $\boldsymbol{\nabla}\overline{\Omega}$ series expansion
contains too many terms for the integration by parts hint to facilitate
obtaining the final relation. An obscure, present day vector analysis
integral theorem provides better direction. The theorem states\begin{equation}
\int\int\int\mathbf{V}_{1}\left(\boldsymbol{\nabla}\bullet\mathbf{V}_{2}\right)dv=\int\mathbf{V}_{1}\mathbf{V}_{2}\bullet d\mathbf{s}-\int\int\int(\mathbf{V}_{2}\bullet\boldsymbol{\nabla})\mathbf{V}_{1}dv.\label{eq:IdThm}\end{equation}
 So, with $\mathbf{V}_{1}=\frac{\mathbf{\mathbf{R}}}{R}$ and $\mathbf{V}_{2}=\mathbf{\mathbf{J}}$\begin{equation}
\left.\begin{array}{ccc}
\int\int\int\frac{\mathbf{\mathbf{R}}}{R}\left(\boldsymbol{\nabla}\bullet\mathbf{\mathbf{J}}\right)dv & = & \int\frac{\mathbf{\mathbf{R}}}{R}\mathbf{\mathbf{J}}\bullet d\mathbf{s}-\int\int\int(\mathbf{\mathbf{J}}\bullet\boldsymbol{\nabla})\frac{\mathbf{\mathbf{R}}}{R}dv,\\
\\\int\int\int(\mathbf{\mathbf{J}}\bullet\boldsymbol{\nabla})\frac{\mathbf{\mathbf{R}}}{R}dv & = & \int\int\int\frac{\mathbf{\mathbf{J}}}{R}dv-\int\int\int\mathbf{\mathbf{J}}\bullet\mathbf{\mathbf{R}}\frac{\mathbf{\mathbf{R}}}{R^{3}}dv.\end{array}\right\} \label{eq:IdThmApp}\end{equation}
 With these Eq.(\ref{eq:L4}) is readily verified when signs are adjusted
for converting the integration and differentiation variables from
$\mathbf{r}$ to $\mathbf{r}'$ in Eq.(\ref{eq:IdThmApp}). The Eq.(\ref{eq:L4})
right hand side is an expansion for which only the first two members
have been retained and includes a negligible term required to restore
retardation to the current density.

\subsection{Retarded Basis}

\subsubsection{Lorenz Formulation\label{sub:Lorenz-formulation}}

If $a$ is assumed to equal the vacuum light speed, the neglected
Eq.(\ref{eq:L4}) terms will be negligible for the small distances
studied, {}``a few feet'', provided the second and higher order
current component time derivatives be {}``not very great''. Under
these conditions Eq.(\ref{eq:L1}) can be rewritten using Eq.(\ref{eq:L4})
to give\begin{equation}
\mathbf{J}=-\kappa\left(\boldsymbol{\nabla}\overline{\Omega}+\frac{2}{c}\frac{\partial\mathbf{A}_{L}}{\partial t}\right).\tag{A}{}\label{eq:L(A)}\end{equation}
Here $\mathbf{A_{L}}$ has the  form\begin{equation}
\mathbf{A}_{L}=\frac{2}{c}\int\int\int\frac{\mathbf{J}'}{R}dv'.\label{eq:AsubL}\end{equation}
Using Eq.(\ref{eq:L4}), Eq.(A) can be written as\begin{equation}
\mathbf{J}=-\kappa\left(\boldsymbol{\nabla}\Omega+\frac{c}{2a^{2}}\frac{\partial\mathbf{A}_{K}}{\partial t}+\left(\frac{4}{c^{2}}-\frac{1}{a^{2}}\right)\frac{c}{2}\frac{\partial\mathbf{A}_{L}}{\partial t}-\cdots\right).\label{eq:(A)var}\end{equation}
Setting $2a=c$ recovers Eq.(\ref{eq:L1}). However, making $a$ {}``infinitely
great'' gives an experimentally confirmed form as well. Thus, $a$
must be regarded as a very great undetermined value for which $a\sqrt{2}=c$
would represent a root mean reciprocal square (RMrS) value for these
extremes.

\subsubsection{Lorenz Formulation Critique\label{sub:Lorenz-Formulation-critique}}

For his constant $c$, Lorenz uses the value\begin{equation}
c_{W}=284736\: miles(/sec).\label{eq:CsubW}\end{equation}
Lorenz attributes $c_{W}$ to experimental determination by Weber
for inductive current transfer. He acknowledges that this value is
about $\sqrt{2}$ times the then known light speed. For the present
day vacuum light speed $c_{0}=299792458\: ms^{-1}$, \begin{equation}
\frac{c_{W}}{c_{0}}=\frac{284736\: miles(/sec)}{299792.458\: kms^{-1}}1.60935\: km/mile=1.529.\label{eq:CsubW/C0}\end{equation}
The $c_{W}$ superluminality is explained in \cite{Mendelson} as
due to a $\sqrt{2}$ scale factor in Weber's 1851 formulaic definition.
Lorenz considers his relation $c_{0}=c_{\text{W}}/\sqrt{2}$ as a
new light speed determination. He, therefore, takes $a=c/\sqrt{2}$.

Recent measurements \cite{Kholmetskii} confirm superluminal near
field magnetic induction propagation speed. There are also several
recent measurements for near field longitudinal electric field propagation
speed with finite values \cite{ScalarPotentialSpeed}. These may indicate
reality for the v-gauge \cite{Jackson} that allows arbitrary scalar
potential propagation speed if other observable effects could be tied
to the gauges. The v-gauge applies to the configuration dependent
potentials which determine the measured fields. These measured fields
can be gauge independent, but as mentioned in \ref{sub:Kirchhoff-Formulation-Critique}
they need not be. Making the Lorenz condition a scalar function has
been proposed as a solution to this impasse \cite{Vlaenderen}. This
solution is supported by the \ref{sub:Lorenz-Condition} Eq.(\ref{eq:RetardedLorenzCondition})
and gives field equations identical to the Lorenz equations presented
in Appendix \ref{sec:Field-Equations}. The Eq.(\ref{eq:OmegaBarZero})
$\overline{\Omega}_{0}$ gives the scalar function a specific physical
definition. To the extent that the v-gauge enters into photonic crystal
\cite{Photonics1,Photonics2} and microwave lens \cite{MicrowaveLens1,MicrowaveLens2}
design, its reality only recently has been subjected to serious study.
Formal field propagation speed expressions are developed in Appendix
\ref{sec:Electromagnetic-Field-Retardation}.

\subsection{Wave Representation}

Having determined that the retardation speed $a$ is not restrained
by the foregoing development, Lorenz proposes to find another method
for its determination and for supporting or correcting that development.
Although he purports to have found several ways that retardation might
be effected, he rejects such hypothecation because its significance
would require selecting a most probable hypothesis. This qualm has
not been an impediment for subsequent authors, however. For examples
see Table \ref{Table:PhotonModels}. Instead, he resolves to further
develop the retardation result with the expectation that added insight
will follow. To this end, he considers wave propagation for a general
source function $\phi$ using the expression\begin{equation}
\square_{a}\int\int\int\frac{\phi\left(t-\frac{R}{a},x',y',z'\right)}{R}dx'dy'dz'=-4\pi\phi\left(t,x,y,z\right)\tag{L5}{}\label{eq:GreenThm}\end{equation}
where, the d'Alembertian, $\square_{a}=\frac{\partial^{2}}{\partial x^{2}}+\frac{\partial^{2}}{\partial y^{2}}+\frac{\partial^{2}}{\partial z^{2}}-\frac{1}{a^{2}}\frac{\partial^{2}}{\partial t{}^{2}}$
.

\subsubsection{Wave formulation\label{sub:Wave-formulation}}

Applying this to Eq.(A) using the Eqs.(\ref{eq:OmegaBar}) and (\ref{eq:AsubL})
definitions gives\begin{equation}
\square_{a}\mathbf{J}=4\pi\kappa\left(\boldsymbol{\nabla}\epsilon'+\frac{4}{c^{2}}\frac{\partial\mathbf{J}}{\partial t}\right)\label{eq:Jwave}\end{equation}
with which the first Eq.(\ref{eq:L2}) relation is associated. Lorenz
displays a one transverse current component, two-dimensional solution
for $\epsilon=0$, periodic in one-dimension and attenuated in the
other. The published solution is inconsistent with the above relations,
because for $\epsilon=0$ at least two current components are required.
However, a suitable one component solution instance is the following:\begin{equation}
\left.\begin{array}{ccc}
J_{x}(x,y,z,t) & = & e^{-\left(z\right)h}\cos\left[p\left(\omega t-z\right)\right]\\
J_{y}(x,y,z,t) & = & 0\\
J_{z}(x,y,z,t) & = & 0\end{array}\right\} \tag{L6\mbox{'}}{}\label{eq:L6sub}\end{equation}
with\begin{equation}
\begin{array}{ccc}
h^{2}a^{2}=p^{2}(a^{2}-\omega^{2}) & \textrm{and} & hc^{2}=8\pi\kappa\omega.\end{array}\tag{L7}{}\label{eq:L7sub}\end{equation}
Both $h$ and $p$ have inverse length units, while $\kappa$ has
inverse time units obtained from and inverse electric resistivity
with F/ms units by electric constant division. As published, Lorenz
had an $x$ substituted for $z$ in the exponential.

For Lorenz this preliminary Eq.(A) treatment demonstrates several
significant results: periodical electrical currents are possible;
these periodical electric currents travel with (phase) velocity, $\omega$;
and they exhibit transverse vibrations, at right angles to the propagation
direction. Furthermore, if light vibrations are assumed to be electric
currents {}`` $a$ is the velocity with which electrical action is
propagated through space.''

\subsubsection{Traveling Wave Critique\label{sub:TravellingWave-Critique}}

The Eq.(\ref{eq:GreenThm}) form presumes $\phi\left(0,x,y,z\right)=0$
with zero first derivatives as initial conditions \cite[pp. 147-148]{Bateman}
and applies only in the region $R<at$ and only up to the nearest
scattering inhomogeneity.

The Eq.(\ref{eq:L6sub}) solution has one critical defect that seems
common to sourced wave equation solutions described as traveling waves:
It does not, and can not, satisfy the initial conditions anywhere
on the solution front at $R=at$. As a consequence, traveling wave
solutions to wave equations for one spatial dimension can not be considered
valid solutions. But for higher spatial dimensions there may be solutions
that incorporate discontinuity at  the wave front \cite[Ch. IX]{Bateman}.
As an alternative ray formulation, Bateman has examined singularity
propagation on wave equation characteristics \cite{BoL}. Being located
\emph{on the expanding wave front}, these singularities provide the
initial condition transition. I propose that such solutions be called
{}``photon solutions'', because with neither wave packet dispersion
\cite{Darwin} nor quantum mechanical non-locality they provide localization
that has long eluded discovery  \cite{Keller}. Bateman acknowledges
that there is no evidence that his singular fields can be built up
by elementary electromagnetic field superposition. He calls these
fields \char`\"{}æthereal\char`\"{}, and proposes that their superposition
defines the electromagnetic field about moving singularities. \cite[Sec. 2]{AetherealFields}

\subsection{Lorenz Condition\label{sub:Lorenz-Condition}}

For use in transforming his Eq.(A) to his Eq.(B) below, Lorenz develops
in the manner next presented what we have come to call the Lorenz
condition.

\subsubsection{Development}

By writing Eqs.(\ref{eq:L2}) with the retarded time $t-\frac{R}{a}$
substituted for the temporal argument and substituting them into Eq.(\ref{eq:OmegaBar})
after $t$ differentiation, he should obtain\begin{equation}
\frac{\partial\overline{\Omega}}{\partial t}=-2\int\int\int\frac{\mathbf{\mathrm{\boldsymbol{\nabla}'\bullet\mathbf{J}'}}}{R}dv'-2\int\frac{\mathbf{J}'\bullet d\mathbf{n}'}{R}.\label{eq:OmegaBarDot}\end{equation}
Lorenz then says integration by parts turns the expression into one
involving the Eq.(\ref{eq:AsubL}) vector potential that we call the
Lorenz condition\begin{equation}
\frac{\partial\overline{\Omega}}{\partial t}=-c\boldsymbol{\nabla}\bullet\mathbf{A}_{L}.\label{eq:LorenzCondition}\end{equation}

\subsubsection{Development Critique\label{sub:Development-Critique}}

I find the transition from Eq.(\ref{eq:OmegaBarDot}) to Eq.(\ref{eq:LorenzCondition})
troubling. Certainly, the explicit surface integration in Eq.(\ref{eq:OmegaBarDot})
can be suppressed to recover a single volume integration, but factoring
out the $\boldsymbol{\nabla}\bullet$ is not obviously valid since
inside the integral it acts on the $\mathbf{r}'$ in $\mathbf{J}(\mathbf{r}',t-\frac{R}{a})$
and outside the integral it acts only on $\mathbf{r}$ in $R$. If
, however, the $\boldsymbol{\nabla}\bullet$ acting on $\mathbf{r}$
and $\boldsymbol{\nabla}'\bullet$ acting on $\mathbf{r}'$ are applied
to the argument in the Eq.(\ref{eq:AsubL}) $\mathbf{A}_{L}$ integral
definition and the results added, we obtain\begin{equation}
-\boldsymbol{\nabla}\bullet\left(\frac{\mathbf{J}'}{R}\right)=\boldsymbol{\nabla}'\bullet\left(\frac{\mathbf{J}'}{R}\right)-\frac{\boldsymbol{\nabla}'\bullet\mathbf{J}'}{R}+\frac{1}{aR^{2}}\mathbf{R}\bullet\frac{\partial\mathbf{J}'}{\partial t}\label{eq:DivVariableShift}\end{equation}
when $\mathbf{J}'=\mathbf{J}(\mathbf{r}',t-\frac{R}{a})$. Significantly,
the temporal derivative from $\boldsymbol{\nabla}\bullet\mathbf{J}'$
remains. It is required to cancel a corresponding term from$\boldsymbol{\nabla}'\bullet\mathbf{J}'$;
but that term can not be explicitly displayed without defining a nonphysical,
derived residue that neglects retardation, \emph{i.e.} by considering
the vector potential to be a non local function whose spatial derivatives
are determined by conditions far from the point at which the derivative
is to be determined. Now, sub\-jecting Eq.(\ref{eq:DivVariableShift})
to volume integration gives the Lorenz condition with retardation\begin{equation}
-c\boldsymbol{\nabla}\bullet\mathbf{A}_{L}=\frac{\partial\overline{\Omega}}{\partial t}+\frac{\partial\overline{\Omega}_{0}}{\partial t}\label{eq:RetardedLorenzCondition}\end{equation}
 when $\int\int\int\boldsymbol{\nabla}'\bullet\left(\frac{\mathbf{J}'}{R}\right)dv'=\int\left(\frac{\mathbf{J}'}{R}\right)\bullet d\mathbf{s}'$,
by the Gauss theorem, vanishes on the volume boundary, and\begin{equation}
\overline{\Omega}_{0}=\frac{2}{a}\int\int\int\frac{\mathbf{R}\bullet\mathbf{J}'}{R^{2}}dv'.\label{eq:OmegaBarZero}\end{equation}
By applying Eq.(\ref{eq:GreenThm}) to Eq.(\ref{eq:RetardedLorenzCondition})
$\overline{\Omega}_{0}$ can be shown to satisfy the homogeneous wave
equation, $\square_{a}\overline{\Omega}_{0}=0.$ Thus, $\overline{\Omega}_{0}$
is a general solution for the configurational scalar potential when
fixed charge is absent. 

This development is unnecessary for showing Eq.(A) and Eq.(B) equivalent
in \ref{sub:Wave-reformulation} below when $\boldsymbol{\nabla}\bullet\mathbf{J}=0$.
However, as Lorenz, himself, confirms in \ref{sub:Wave-reformulation}
$\boldsymbol{\nabla}\bullet\mathbf{J}=0$ is the necessary and sufficient
condition for replacing electromagnetic action at a distance, Eq.(A),
by strict locality, Eq.(B). This means Eq.(\ref{eq:OmegaBarDot})
is identically zero. So the Eq.(\ref{eq:LorenzCondition}) Lorenz
condition reduces to the Coulomb gauge, $\boldsymbol{\nabla}\bullet\mathbf{A}_{L}=0$;
but Eq.(\ref{eq:RetardedLorenzCondition}) shows this to be true only
when the Eq.(\ref{eq:OmegaBarZero}) temporal derivative also vanishes.
So the correction to the Eq.(\ref{eq:LorenzCondition}) Lorenz condition
found here has nontrivial consequences. Some are expounded in the
appendices. Importantly, the Eq.(\ref{eq:RetardedLorenzCondition})
Lorenz condition with retardation gives modified field equations developed
in Appendix \ref{sec:Field-Equations}. Maxwell equation dependence
on the Lorenz condition has been examined recently \cite{Onoochin,Vlaenderen}.

\subsection{Light Equation\label{sub:LightEquation}}

\subsubsection{Wave Reformulation\label{sub:Wave-reformulation}}

Through a complex mathematical manipulation that we would effect by
using the vector identity\begin{equation}
\boldsymbol{\nabla}\times\left(\boldsymbol{\nabla}\times\mathbf{J}\right)=\boldsymbol{\nabla}\left(\boldsymbol{\nabla}\bullet\mathbf{J}\right)-\nabla^{2}\mathbf{J}\label{eq:CcThm}\end{equation}
in Eq.(\ref{eq:Jwave}) for charge free space, Lorenz succeeds in
rewriting his Eq.(\ref{eq:L(A)}). With his $c=a\sqrt{2}$ the result
is his light equation augmented with a conductivity term that introduces
attenuation\begin{equation}
\boldsymbol{\nabla}\times\left(\boldsymbol{\nabla}\times\mathbf{J}\right)+\frac{1}{a^{2}}\frac{\partial^{2}\mathbf{J}}{\partial t^{2}}+8\pi\kappa\frac{1}{a^{2}}\frac{\partial\mathbf{J}}{\partial t}=0.\tag{B}{}\label{eq:L(B)}\end{equation}
He also shows that the original equations can be recovered through
another mathematical manipulation made more complex than the first
by the necessity to reintroduce the non-zero charge density gradient.
These manipulations are more fully described in Appendix \ref{sec:A_Bequivalence}.
During both he explicitly displays the vector potential wave function
that follows from the Eq.(\ref{eq:GreenThm}) paradigm applied to
Eq.(\ref{eq:AsubL}): \begin{equation}
\square_{a}\mathbf{A}_{L}=-\frac{8\pi}{c}\mathbf{J}.\label{eq:AsubLwave}\end{equation}

The vibrations described by Eqs.(\ref{eq:L(B)}) are transverse; longitudinal
vibrations will not be possible. The first is confirmed in \ref{sub:Wave-formulation};
the second, because for $\boldsymbol{\nabla\mathrm{\bullet\mathbf{J}}}\equiv\Theta$,
$\boldsymbol{\nabla\mathrm{\bullet}}$ Eq.(\ref{eq:L(B)}) gives\begin{equation}
\frac{\partial\Theta}{\partial t}+8\pi\kappa\Theta=0\label{eq:DivTheta}\end{equation}
after one time integration. Clearly $\Theta$ can not be periodic
and $\lim_{t\rightarrow\infty}\Theta=0$. Since Eqs.(\ref{eq:L2})
give $\Theta=-\frac{1}{2}\frac{\partial\epsilon}{\partial t}$, no
free charge {}``development'' is possible within a conductor.

\subsubsection{Wave Reformulation Critique\label{sub:Wave-Reformulation-Critique}}

As discussed more fully in Appendix \ref{sec:Field-Equations}, Lorenz
claims to have solved the self field, action at a distance problem
inherent in Eq.(\ref{eq:L(A)}) with $\mathbf{A}_{L}$ defined by
Eq.(\ref{eq:AsubL}). The Eq.(\ref{eq:L(B)}) which contains only
one, local field is his solution. The Appendix \ref{sec:Field-Equations}
also shows that Eq.(\ref{eq:L(B)}) is not required to obtain the
field equations from the foregoing Lorenz development when Eq.(\ref{eq:Bdef})
is used to define the magnetic induction. Lorenz, himself, never gives
an expression for the magnetic field. He only mentions that its propagation
speed is apparently superluminal as covered in \ref{sub:Lorenz-Formulation-critique}.
Electromagnetic field retardation is discussed more fully in Appendix
\ref{sec:Electromagnetic-Field-Retardation}.

\subsection{Concluding Remarks}

Lorenz ends the work with paragraphs giving a summary, a suggestion
for extending the development to heterogeneous material, an æther
dismissal and, of course, his alternative conception for light vibration.

\subsubsection{Lorenz Conclusion}

Lorenz considers the work to be a {}``new proof'' that light vibrations
are electrical currents. He claims to have shown that vibrations described
by Eq.(\ref{eq:L(B)}), his {}``laws of light'' augmented by absorption,
are electric currents described by Eq.(\ref{eq:L(A)}). This result
is said to depend on no assumed physical hypothesis.

Lorenz deems extension to heterogeneous bodies best done by starting
from the differential equations, possibly an indirect reference to
the Maxwell relations given in \ref{sub:Kirchhoff-Formulation-Critique},
while regarding $a$ and $\kappa$ to be variable quantities. Lorenz
laments that the Eq.(\ref{eq:L(A)}) simplicity could not be preserved.
So Eq.(\ref{eq:L(A)}) must be considered as strictly applying to
homogeneous bodies. The important result, however, is that {}``electrical
forces require time to travel.'' As a result, the action at a distance,
non-locality, contained in Eq.(\ref{eq:L(A)}) in fact results from
local action described by Eq.(\ref{eq:L(B)}).

Since the æther theory for light at the time was one with æther particles
in motion, the Lorenz conception that light vibrations were electric
currents implied æther particles to be electric in nature. So electric
currents would be described by progressive æther particle motion.
Lorenz found this prospect untenable. So, rather than reject his electric
current association with light vibrations, he found it expedient to
reject the æther concept.

The linearly transverse current solution given by Eq.(\ref{eq:L6sub})
gives Lorenz no reason to prefer his electric current model for light
vibrations over the æther model. But he justifies the preference by
proposing that the currents are actually rotary. In good conductors
current circulates in the rotation direction, but in bad conductors
it is propagated by induction in a direction perpendicular to the
planes containing the periodical currents.

\subsubsection{Conclusion Critique\label{par:Rotary-Current-Critique}}

For Lorenz, the conception that electrical interactions require time
for propagation is so obvious and scientifically acceptable that it
may be assumed as fact. With this assumed physical hypothesis, Lorenz
has bypassed the Maxwell field formulation that we consider fundamental.

Although Lorenz states that he had previously {}``adverted'' to
the rotary light coupling, he presents no explicit example in his
paper. However, the proposal can be looked upon as magnetoinductive
coupling. This effect has been demonstrated recently to transfer
60 watts with 40\% efficiency between self-resonant coils with resonant
frequency near 10 MHz over a distance exceeding 2 meters in a strongly
coupled regime \cite{PowerTransfer}. This power transfer does not
have simple Poynting character. If the energy were to propagate by
inductive regeneration as proposed by Lorenz, the rotary electric
field would progress autonomously with alternating electric field
direction. This possibility might be experimentally evaluated in the
wireless power transfer apparatus  by attempting to measure the induced
electromotive force phase and power transfer efficiency change produced
by adding one or more resonant coils between the source and receiver
coils. These waves add to self-trapping \cite{Piekara} and synchrotron
electric fields \cite[Figure 5]{Hanney} as possible Maxwell wave
alternatives.

For photons Dirac \cite{Dirac} has derived a phase space volume equal
to the Planck constant cubed. With momentum equipartition this gives
an effective spatial volume per photon equal to $\left[3\lambda\right]^{3}$,
each containing the photon energy $h\nu$. Magnetosphere continuous
pulsations at 0.005 Hz investigated by satellite \cite{PhaseSkip}
have revealed antiphase electric and magnetic fields within the theoretical
spatial volume equal to 1.1x10$^{\text{24}}$ km$^{\text{3}}$, about
10$^{\text{6}}$ sun volumes. With a linear dimension equal to 10$^{\text{8}}$
km, passage at vacuum light speed in 5 minutes or less agrees well
with the reported pulse durations. If these observations can be attributed
to individual photons, a new study regime may be opened.

\section{Discussion}

Frequently in attempting to solve complex problems, we find ourselves
retracing paths that fail to provide an adequate solution. Table \ref{Table:PhotonModels}
shows that our search for a physical photon model is an example. In
such cases, a fresh start can often help to break free from the limitations
that impede progress. The fresh start can generally be either closely
examining prior attempts for possible flaws or taking an entirely
different approach. In subjecting the 1867 Lorenz paper to detailed,
critical examination, this paper applies both these remedies for failed
attempts to resolve light wave/particle duality.

Table \ref{Table:PhotonModels} examination shows, and \cite{Keller}
confirms, that nearly all attempts to describe particle like light
behavior are based on the Maxwell electromagnetism formulation. Being
written before Maxwell attained unquestioned authority, the 1867 Lorenz
paper begins at the well established Ohm's law level as formulated
by Kirchhoff, Eq.(\ref{eq:L1}), with the electric field defined by
two potentials, a scalar potential and a vector potential. By introducing
retardation, apparently a novel idea at the time, to eliminate potential
tracking instantaneously with source position \cite{Smirnov-Rueda},
Lorenz obtains his Eq.(A) Ohm's law formulation. Using Eq.(\ref{eq:(A)var})
Lorenz demonstrates that the retardation speed is not theoretically
established. By choosing $a=c_{\text{0}}=c_{\text{W}}/\sqrt{2}$,
however, he interrelates the propagation speeds for inductive current
transfer and for light. This choice for scalar potential propagation
speed was experimentally validated only recently \cite[Tsontchev]{ScalarPotentialSpeed}.

To obtain more insight into electric signal propagation, Lorenz applies
the Green's function solution for the sourced wave equation, Eq.(\ref{eq:GreenThm}),
to his retarded potential Ohm's law formulation, Eq.(\ref{eq:L(A)}).
The result, Eq.(\ref{eq:Jwave}), is effectively a sourced wave equation
for the electric field. Lorenz presents a solution for this equation
in charge free space. After typographical error correction, the solution
represents a plane polarized, traveling wave exhibiting light like
properties by being periodic and transverse to the propagation direction.
In \ref{sub:TravellingWave-Critique} I note that traveling waves
do not satisfy initial conditions on the expanding solution boundary.
But there appear to be singular wave equation solutions that can satisfy
the initial conditions on the expanding wave front \cite{BoL}. Such
solutions may represent the elusive localized photon.

For use in transforming Eq.(\ref{eq:Jwave}) into a form that he had
previously published as an equation for light, Lorenz develops what
we call the Lorenz condition, Eq.(\ref{eq:LorenzCondition}), from
Eq.(\ref{eq:OmegaBarDot}). Flaws are exposed in \ref{sub:Lorenz-Condition}
and \ref{sub:LightEquation}. The \ref{sub:Lorenz-Condition} development
must be corrected to avoid inadvertent retardation suppression in
the vector potential current density argument. This correction yields
field equations derived in Appendix \ref{sec:Field-Equations} with
an augmented Maxwell form. Furthermore, in the \ref{sub:LightEquation}
development, he then shows that $\boldsymbol{\nabla}\bullet\mathbf{J}=0$.
So, from Eq.(\ref{eq:OmegaBarDot}) the scalar potential temporal
time derivative vanishes and the Lorenz condition without correction,
Eq.(\ref{eq:LorenzCondition}), reduces to $\boldsymbol{\nabla}\bullet\mathbf{A}_{\text{L}}=0$,
the Coulomb gauge.

Only recently has the magnetoinductive wave light model proposed by
Lorenz been subjected to observational \cite{Wiltshire} and theoretical
analysis \cite{Syms}. These models are formed from inductively coupled,
capacitively loaded wire rings arranged in 1, 2 and 3 dimensional
arrays to simulate common optical phenomena as multiport networks.
These elements can not, however, be described by the Lorenz Eq.(\ref{eq:L(B)}),
because the capacitive loading represents free charge development.
We do not yet know if the Appendix \ref{sec:Field-Equations} Lorenz
fields will support magnetoinductive waves, but the requisite structure
may be plasma emergent \cite[VI.]{Kinney}.  More conventionally,
however, the capacitor electric field could be taken to describe a
wave polarization orientation that varies with capacitor azimuth.
Considered as a series connected LC loop, each module would exhibit
a representative capacitor charge, $Q_{C}=CV_{0}\cos\omega t$, where
$\omega^{2}=\left(LC\right)^{-1}$ for initial capacitor voltage $V_{0}$
and loop inductance $L$. The loop current, $i_{C}=-\omega V_{0}\sin\omega t$,
would produce a time dependent magnetic field through the loop coupling
it to its near neighbors with some delay. The electric field defined
by the capacitor voltage, $V_{C}=Q_{C}/C=V_{0}\cos\omega t$, clearly
differs in phase from the developed magnetic field, proportional to
$i_{C}$, by $\pm\frac{\pi}{2}$. To describe general electromagnetic
phenomena, these modules would have to be vectorized to properly represent
their azimuthal and axial orientation. With this and the axial magnetic
field, the modules can be linearly configured to display  wave modes
presented in Table \ref{tab:Magnetoinductive-Wave-Modes}. The Kirchhoff
network waves displayed in Table \ref{tab:Magnetoinductive-Wave-Modes}
exhibit modes that are generally excluded from consideration for Maxwell
waves and represent a proper complement to the Lorenz legacy.

\section{Conclusion}

The Lorenz 1867 paper has both blemishes and exceptional innovations.
The most obvious and disconcerting blemish is the typographical error
in Eq.(\ref{eq:L6sub}). The erroneous solution to his Eqs.(\ref{eq:Jwave})
casts doubt on the whole paper even though the minor revision provided
in Eq.(\ref{eq:L6sub}) does work. By modern standards his equating
light vibrations to electrical currents rather than electric fields
is archaic, but future magnetoinductive wave studies may well prove
his usage prescient. His Eq.(\ref{eq:L(B)}) development from Eq.(\ref{eq:L(A)})
using the effectively null relation that we associate with the Lorenz
condition, Eq.(\ref{eq:LorenzCondition}), rather than proceeding
directly from his Eqs.(\ref{eq:Jwave}) is especially egregious. Most
likely he did not want to explicitly set the fixed charge in Eqs.(\ref{eq:Jwave})
to zero, but found he could do it implicitly with Eq.(\ref{eq:LorenzCondition}).
The \ref{sub:Development-Critique} Eq.(\ref{eq:RetardedLorenzCondition})
retarded Lorenz condition reveals a heretofore unrecognized scalar
potential,  $\overline{\Omega}_{0}$, developed by currents, Eq.(\ref{eq:OmegaBarZero}).
The most pernicious problem, one that persists to this day, is taking
the traveling wave solution to the wave equation as a complete, valid
solution for wave propagation. This solution by itself does not satisfy
the initial conditions on the expanding wave front and is not, therefore,
a complete, valid solution. In his book \cite{Bateman}, Bateman gives
this problem serious consideration, but his work is unfinished and
largely ignored as incoherent \cite{BatemanReview}. A major contribution
to our science could result from continuing his work.

The most innovative Lorenz contribution is effective Maxwell equation
development, see \ref{sub:Kirchhoff-Formulation-Critique} and Appendix
\ref{sec:Field-Equations}, from Kirchhoff's expression for Ohm's
law by introducing scalar potential retardation. Although not an astounding
innovation, the retained source terms and real value functions with
retarded time arguments are exemplary. Now days we commonly suppress
sources and use traveling wave phase without considering how the far
field waves are to emerge from the retarded near fields. Another modern
day time saver uses the complex exponential for periodic functions.
By reducing time dependence to a multiplicative factor, this expedient
conceals the emergence problem by a subtle change in spatial argument
from $\frac{R}{a}$ to $\frac{\mathbf{k\bullet R}}{\omega}$ and blatantly
confounds Helmholtz Theorem components with longitudinal and transverse
fields \cite[Sec. 3.]{Keller}. The magnetoinductive wave discussion
with reference to Table \ref{tab:Magnetoinductive-Wave-Modes} shows
these expedients can mask real phase and orientation relationships.
In starting from Ohm's law, Lorenz has implicitly overcome the long
standing myopic focus on {}``visible light.'' Extending over a 10$^{\text{30}}$
factor for frequency and wavelength, Table \ref{Table:PhotonCharacteristics}
displays only a minute range for exploration. The lowest, \cite{PhaseSkip},
and highest, \cite{DeRoeck}, are science frontiers. The first may
indicate that studying photons from the inside may actually be possible
as indicated in \ref{par:Rotary-Current-Critique}.

Lorenz most prescient contribution may be his paper ending suggestion
that light is rotationally polarized, \emph{i.e.} the electric field
is azimuthally directed in cylindrical coordinates rather than radially
directed, and propagates by induction along the cylinder axis. This
suggestion I have interpreted to mean light actually has a magnetoinductive
wave character. Lorenz could not properly develop this concept from
his most favored Eq.(\ref{eq:L(B)}) light wave representation, because
magnetoinduction appears to require the free charge in Eqs.(\ref{eq:Jwave})
and, thus, Eq.(\ref{eq:L(A)}). Whether such waves are possible with
the Appendix \ref{sec:Field-Equations} Lorenz fields and Bateman
localization is an important subject for future study. But, now, after
more than a century, there are signs \cite{Syms} that magnetoinduction
may provide a tool for advancing our long standing bond to Maxwell.

The $\overline{\Omega}_{0}$ term in \ref{sub:Development-Critique}
comes from a by parts integration in which the vector potential is
treated as a local function with spatial derivatives at a point determined
by proximate function values. In reviewing this paper I found that
the Eq.(\ref{eq:LorenzCondition}) Lorenz condition can be developed
from the non local vector potential. This follows from applying $\boldsymbol{\nabla}\bullet$
to Eq.(\ref{eq:AsubLwave}) and using the Eq.(\ref{eq:L2}) charge
continuity condition to obtain\begin{equation}
\square_{a}\boldsymbol{\nabla}\bullet\mathbf{A}_{L}=-\frac{8\pi}{c}\boldsymbol{\nabla}\bullet\mathbf{J}=\frac{4\pi}{c}\frac{\partial\epsilon}{\partial t}.\label{eq:DivAsubLwave}\end{equation}
So applying Eq.(\ref{eq:GreenThm}) to Eq.(\ref{eq:OmegaBar}) gives
Eq.(\ref{eq:LorenzCondition}) to within an additive homogeneous wave
equation solution. Thus, the Lorenz condition to be a charge continuity
condition restatement for wave function potentials. It allows the
field equations to be instructively developed when the potentials
treated consistently \cite{ELF}. In this development, the homogeneous
wave function gives nongauge polarization fields. For potentials with
independent plane wave vector components, these can emerge from retarded
Maxwell fields as light like rays that propagate autonomously by conserving
charge. Alternative treatments using separated-potentials are listed
in \cite{Chub&S-R}.

\section{Acknowledgments}

I would like to thank the Louisville Free Public Library interlibrary
loan staff for their valiant efforts in obtaining many references
that have made this work possible and especially Teresa Frost whose
exceptional facility management helped to provide a felicitous work
environment. I would also like to thank an anonymous American Journal
of Physics reviewer for suggesting \cite{BatemanReview}.

\appendix

\section{Eq.(\ref{eq:L(A)}) and Eq.(\ref{eq:L(B)}) Equivalence\label{sec:A_Bequivalence}}

As mentioned in \ref{sub:Wave-reformulation}, Lorenz presents arguments
to show that his Eq.(\ref{eq:L(A)}) and Eq.(\ref{eq:L(B)}) are equivalent.
The argument from Eq.(\ref{eq:L(A)}) to Eq.(\ref{eq:L(B)}) skirts
the intermediate Eq.(\ref{eq:Jwave}). If, however, the vector identity
Eq.(\ref{eq:CcThm}) is applied to Eq.(\ref{eq:Jwave}), Eq.(\ref{eq:L(B)})
is equivalent to Eq.(\ref{eq:Jwave}) and, therefore, to Eq.(\ref{eq:L(A)})
if and only if $\boldsymbol{\nabla}(\boldsymbol{\nabla}\bullet\mathbf{J})=4\pi\kappa\boldsymbol{\nabla}\epsilon'$.
Thus, using Eq.(\ref{eq:L2}), equivalence demands\begin{equation}
\frac{\partial\boldsymbol{\nabla}\epsilon'}{\partial t}+8\pi\kappa\boldsymbol{\nabla}\epsilon'=0.\label{eq:Grade}\end{equation}
This complements the related Eq.(\ref{eq:DivTheta}) for $\boldsymbol{\nabla}\bullet\mathbf{J}$.
Apparently, Lorenz has demonstrated Eq.(\ref{eq:Grade}) validity
without ever explicitly stating it. To see how, I here recapitulate
the Lorenz argument for obtaining Eq.(\ref{eq:L(B)}) from Eq.(\ref{eq:L(A)}).

From Eq.(\ref{eq:L(A)})\begin{equation}
\frac{-1}{\kappa}\frac{\partial\mathbf{J}}{\partial t}=\boldsymbol{\nabla}\frac{\partial\overline{\Omega}}{\partial t}+\frac{2}{c}\frac{\partial^{2}\mathbf{A}_{L}}{\partial t^{2}}.\label{eq:LA'}\end{equation}
Using Eq.(\ref{eq:AsubLwave}) with $a\sqrt{2}=c,$ this becomes\begin{equation}
\frac{1}{2\kappa}\frac{\partial\mathbf{J}}{\partial t}+4\pi\mathbf{J}=\frac{-1}{2}\left(\boldsymbol{\nabla}\frac{\partial\overline{\Omega}}{\partial t}+c\nabla^{2}\mathbf{A}_{L}\right).\label{eq:L8'}\end{equation}
Applying the vector identity Eq.(\ref{eq:CcThm}) gives\begin{equation}
\frac{1}{2\kappa}\frac{\partial\mathbf{J}}{\partial t}+4\pi\mathbf{J}=\frac{a}{\sqrt{2}}\boldsymbol{\nabla}\times\left(\boldsymbol{\nabla}\times\mathbf{A}_{L}\right)-\frac{1}{2}\boldsymbol{\nabla}\left(\frac{\partial\overline{\Omega}}{\partial t}+c\boldsymbol{\nabla}\bullet\mathbf{A}_{L}\right).\tag{L8}{}\label{eq:L8}\end{equation}
Going back to Eq.(\ref{eq:L(A)}), Lorenz then notes that\begin{equation}
\boldsymbol{\nabla}\times\mathbf{J}=\frac{-\sqrt{2}\kappa}{a}\frac{\partial\boldsymbol{\nabla}\times\mathbf{A}_{L}}{\partial t}.\tag{L9}{}\label{eq:L9}\end{equation}
Combining $\frac{\partial}{\partial t}$ Eq.(\ref{eq:L8}) with $\boldsymbol{\nabla}\times$
Eq.(\ref{eq:L9}) to eliminate $\mathbf{A}_{L}$ gives Eq.(\ref{eq:L(B)})
when the Eq.(\ref{eq:LorenzCondition}) Lorenz condition is used.
But clearly, Eq.(\ref{eq:L8}) retains a $\frac{\partial\overline{\Omega_{0}}}{\partial t}$
term when the Eq.(\ref{eq:RetardedLorenzCondition}) Lorenz condition
with retardation is used. By Eq.(\ref{eq:DivTheta}), forcing this
residual term to zero effectively forces the free charge development
to zero. As this results from using the Eq.(\ref{eq:LorenzCondition})
in Eq.(\ref{eq:L8}), the Lorenz condition must be erroneous when
the vector potential is considered to be a function with spatial derivatives
determined by function values in a small region about the point where
the derivative is being determined. The reason is retardation neglect
in the by parts integration for its development as discussed in \ref{sub:Lorenz-Condition}.

To show their equivalence, Lorenz also demonstrates that Eq.(\ref{eq:L(A)})
can be obtained from Eq.(\ref{eq:L(B)}) when Eq.(\ref{eq:L(B)})
is taken to describe the retarded potential field about some finite
volume within which the current actually resides, \emph{i.e.} in Eq.(\ref{eq:L(B)})
$\mathbf{A}_{L}$ can replace $\mathbf{J}$! Lorenz justifies this
substitution term wise by repeated by parts integration over the conduction
volume with zero interface current. When Eq.(\ref{eq:L(B)}) has been
rewritten with $\mathbf{A}_{L}$ substituted for $\mathbf{J}$, Eqs.(\ref{eq:AsubLwave})
and (\ref{eq:LorenzCondition}) can be used to obtain the form\begin{equation}
\frac{1}{2}\boldsymbol{\nabla}\frac{\partial\overline{\Omega}}{\partial t}=4\pi\mathbf{J}+\frac{4\sqrt{2}\pi\kappa}{a}\mathbf{A}_{L}.\label{eq:L(BA)int}\end{equation}
Eliminating $\overline{\Omega}$ from Eq.(\ref{eq:L(BA)int}) by pairs
yields Eq.(\ref{eq:L9}) for $\boldsymbol{\nabla}\times\mathbf{J}$.
Now, substituting Eq.(\ref{eq:L9}) into Eq.(\ref{eq:L(B)}) and integrating
temporally gives Eq.(\ref{eq:L8}) from which Eq.(\ref{eq:LA'}) can
be obtained by reversing the process from which Eq.(\ref{eq:L8})
was obtained from Eq.(\ref{eq:LA'}). One last temporal integration
then recovers Eq.(\ref{eq:L(A)}).

Biologically, the above Lorenz developments may be looked upon as
hermaph\-roditic or even incestuous, because they exhibit unusual
recursion. This characteristic alone makes their validity suspect.
This suspicion is confirmed in Appendix \ref{sec:Field-Equations}.
Fortunately, although expressing less physical configuration dependence
the modern day vector analysis formalism does validate the final result
with the specific restriction that $\boldsymbol{\nabla}\bullet\mathbf{J}=0$.
So in addition to showing how mathematical advances can enhance insight,
this example may provide a case study for evaluating proof theory.

\section{Field Equations\label{sec:Field-Equations}}

As stated in \ref{sub:Kirchhoff-Formulation-Critique} the Lorenz
development appears to be consistent with the Maxwell equations when
these equations are taken to be strictly local. On this basis, then,
the Maxwell equations are simply descriptive, not predictive. To be
predictive the equations must, per Lorenz, reflect retardation. In
his 1867 paper, Lorenz considers scalar potential retardation as expressed
by Eq.(\ref{eq:OmegaBar}). From this he obtains the retarded, local
Eq.(\ref{eq:L1}) electric field as presented in \ref{sub:Retardation}
Eq.(\ref{eq:L4}) for $2a=c$. His Eq.(\ref{eq:L(A)}) predicts the
local current field determined by the retarded scalar and vector potentials
with the latter given by Eq.(\ref{eq:AsubL}). Lorenz never mentions
the magnetic field. Per \ref{sub:Kirchhoff-Formulation-Critique},
the Faraday law for local fields would require the local magnetic
field be given by Eq.(\ref{eq:Bdef}).

Lorenz considers the integral definitions for the retarded potentials
to be Green's function solutions for Eq.(\ref{eq:GreenThm}) wave
equations. Properly, these have vanishing values and first temporal
derivatives on the expanding solution front.  Lorenz solves the self-field
or action at a distance problem by eliminating all but one field from
his mathematical expressions. In particular, he rewrites his Eq.(\ref{eq:L(A)}),
which interrelates $\mathbf{J}$, $\overline{\Omega}$ and $\mathbf{A}_{L}$,
as his Eq.(\ref{eq:L(B)}), which involves only $\mathbf{J}$. If
this were valid, then Eq.(\ref{eq:L(A)}) could be written using only
$\overline{\Omega}$ or $\mathbf{A}_{L}$ as well. This can be effected
for $\mathbf{A}_{L}$ by using the Eq.(\ref{eq:LorenzCondition})
Lorenz condition and Eq.(\ref{eq:AsubLwave}), but by the Eq.(\ref{eq:GreenThm})
Green's function paradigm $\mathbf{A}_{L}$ is nonlocal. This confirms
the suspicion expressed in Appendix \ref{sec:A_Bequivalence} that
replacing $\mathbf{J}$ in Eq.(\ref{eq:L(B)}) with $\mathbf{A}_{L}$
might not be valid.

Since the single function differential relations for $\overline{\Omega}$
and $\mathbf{A}_{L}$ are intractable, their integral definitions,
Eqs.(\ref{eq:OmegaBar}) and (\ref{eq:AsubL}), along with the augmented
Lorenz condition, Eq.(\ref{eq:RetardedLorenzCondition}), and Eq.(\ref{eq:L(A)})
defining $\mathbf{J}$ and $\mathbf{E}$ and Eq.(\ref{eq:Bdef}) defining
$\mathbf{\mathbf{B}}$ define the Lorenz electromagnetic theory. The
retarded Maxwell equations can be derived from these:\begin{equation}
\boldsymbol{\nabla}\bullet\mathbf{\mathbf{B}}=\frac{2}{c}\boldsymbol{\nabla}\bullet(\mathbf{\boldsymbol{\nabla}}\times\mathbf{A}_{L})=0;\tag{M1}{}\label{eq:M1B}\end{equation}
\begin{equation}
\boldsymbol{\nabla}\bullet\mathbf{E}=-\boldsymbol{\nabla}\bullet\left(\boldsymbol{\nabla}\overline{\Omega}+\frac{2}{c}\frac{\partial\mathbf{A}_{L}}{\partial t}\right)=4\pi\epsilon'+\left(\frac{2}{c^{2}}-\frac{1}{a^{2}}\right)\frac{\partial^{2}\overline{\Omega}}{\partial t{}^{2}}+\frac{2}{c^{2}}\frac{\partial^{2}\overline{\Omega}_{0}}{\partial t{}^{2}};\tag{M2}{}\label{eq:M2E}\end{equation}
\begin{equation}
\mathbf{\boldsymbol{\nabla}}\times\mathbf{E}=-\frac{\partial\mathbf{\mathbf{B}}}{\partial t};\tag{M3}{}\label{eq:M3F}\end{equation}
$\begin{array}{cccc}
\mathbf{\boldsymbol{\nabla}}\times\mathbf{B} & = & \frac{2}{c}\boldsymbol{\nabla}\times(\mathbf{\boldsymbol{\nabla}}\times\mathbf{A}_{L}) & =\end{array}$$ $\begin{equation}
=\frac{2}{c^{2}}\left[\frac{\partial\mathbf{E}}{\partial t}+8\pi\mathbf{J}+c\left(\frac{2}{c^{2}}-\frac{1}{a^{2}}\right)\frac{\partial^{2}\mathbf{A}_{L}}{\partial t{}^{2}}-\frac{\partial\boldsymbol{\nabla}\overline{\Omega}_{0}}{\partial t}\right]\tag{M4}{}.\label{eq:M4A}\end{equation}
The Eq.(\ref{eq:M4A}) development first uses Eq.(\ref{eq:CcThm})
and then Eqs.(\ref{eq:RetardedLorenzCondition}) and (\ref{eq:AsubLwave})
before finally using the Eq.(\ref{eq:L(A)}). These simplify considerably
with the $c=a\sqrt{2}$ Lorenz assignment. But even then Eqs.(\ref{eq:M2E})
and ( \ref{eq:M4A}) retain a residual, nonlocal Eq.(\ref{eq:OmegaBarZero})
$\overline{\Omega}_{0}$ dependence. Like the field equations for
moving charges that supplement the conventional Maxwell displacement
current with a dielectric convection current \cite{Chub&S-R2}, the
Eqs. (\ref{eq:M1B})-(\ref{eq:M4A}) have the displacement current
supplemented by a term due to $\overline{\Omega}_{0}$ in Eq.(\ref{eq:M4A}).
But $\overline{\Omega}_{0}$ also modifies the charge density in Eq.(\ref{eq:M2E}).

\section{Electromagnetic Field Retardation\label{sec:Electromagnetic-Field-Retardation}}

In his 1867 paper, Lorenz introduced scalar potential retardation
to obtain his Eq.(\ref{eq:L(A)}) for the  current density $\mathbf{J}$,
effectively the electric field $\mathbf{E}$. Eq.(\ref{eq:L(A)})
has the same form as Eq.(\ref{eq:L1}), but now depends on local,
retarded scalar and vector potentials given by Eqs.(\ref{eq:OmegaBar})
and (\ref{eq:AsubL}). Since $\mathbf{A}_{L}$ is derived from $\mathbf{J}$,
however, Eq.(\ref{eq:L(A)}) has the self-field, action at a distance,
problem. Lorenz resolves this problem with his Eq.(\ref{eq:L(B)})
that depends only on $\mathbf{J}$. So Eq.(\ref{eq:L(B)}) can be
taken to describe electric field propagation. Eq.(\ref{eq:L(B)})
is not a wave equation and comes with the condition that $\boldsymbol{\nabla}\bullet\mathbf{J}=0$,
or $\boldsymbol{\nabla}\bullet\mathbf{E}=0$ when $\kappa\neq0$.
For consistency, then, the $\boldsymbol{\nabla}\bullet\mathbf{E}=0$
condition must be retained when $\kappa=0$. Thus, the electric field
propagation speed will be $a$ and the $\boldsymbol{\nabla}\bullet\mathbf{E}=0$
condition will assure there is no electric field in the propagation
direction, no longitudinal electric field.

The Lorenz field equations in Appendix \ref{sec:Field-Equations}
do not have the $\boldsymbol{\nabla}\bullet\mathbf{J}=0$ restriction.
From $\mathbf{\boldsymbol{\nabla}}\times$ Eq.(\ref{eq:M3F}) and
Eq.(\ref{eq:M4A}) the Eq.(\ref{eq:L(B)}) equivalent is found to
be \begin{equation}
\boldsymbol{\nabla}\times\left(\boldsymbol{\nabla}\times\mathbf{E}\right)+\frac{2}{c^{2}}\frac{\partial^{2}(\mathbf{E}-\boldsymbol{\nabla}\overline{\Omega}_{0})}{\partial t^{2}}+\frac{16\pi}{c^{2}}\frac{\partial\mathbf{J}}{\partial t}=0\label{eq:LM(B')}\end{equation}
when the higher order temporal derivatives are neglected as possibly
having a zero coefficient. If $\mathbf{E}-\boldsymbol{\nabla}\overline{\Omega}_{0}$
in Eq.(\ref{eq:LM(B')}) were mistakenly taken as $\mathbf{E}$, formally
the effective propagation speed would satisfy the relation \begin{equation}
C_{\mathbf{E}}^{2}\left|\frac{\partial^{2}(\mathbf{E}-\boldsymbol{\nabla}\overline{\Omega}_{0})}{\partial t^{2}}\right|=c^{2}\left|\frac{\partial^{2}\mathbf{E}}{\partial t^{2}}\right|.\label{eq:Eceff}\end{equation}
In principle $C_{\mathbf{E}}^{2}$ would be direction and location
dependent and could be greater than $c^{2}$. This argument for a
possible superluminal propagation speed cannot be extended to $\boldsymbol{\nabla}\times\left(\boldsymbol{\nabla}\times\mathbf{B}\right)$,
 because the $\boldsymbol{\nabla}\overline{\Omega}_{0}$ in Eq.(\ref{eq:M4A})
drops out. But the argument does apply to $\boldsymbol{\nabla}\times\mathbf{B}$,
itself, to give\begin{equation}
C_{\boldsymbol{\nabla}\times\mathbf{B}}^{2}\left|\frac{\partial(\mathbf{E}-\boldsymbol{\nabla}\overline{\Omega}_{0})}{\partial t}\right|=c^{2}\left|\frac{\partial\mathbf{E}}{\partial t}\right|.\label{eq:Bceff}\end{equation}

\newpage{}

\begin{center}%
\begin{table}

\caption{Photon Models\label{Table:PhotonModels}}

The table entries are keys to appropriate references.

\begin{centering}\begin{tabular}{ccc}
\hline 
Author&
Date&
Description\tabularnewline
\hline
\hline 
Fermat&
1601-65&
Space-time geodesic\tabularnewline
J. C. Maxwell&
1865+&
Field equations\tabularnewline
L. Lorenz \cite{LLorenz}&
1867&
Inductive, autonomous progression\tabularnewline
H. Bateman \cite{Bateman}&
1915$\pm$&
Singular wave functions\tabularnewline
A. Einstein&
1916&
Nadel Strahlung\tabularnewline
H. Bateman&
1921&
Electric doublet\tabularnewline
G. N. Lewis \cite{Lewis_Light}&
1926&
Relativistic proper-time reality\tabularnewline
P. A. M. Dirac \cite{Dirac}&
1958&
Second Quantization and Quantum Electrodynamics\tabularnewline
W. H. Bostick&
1961&
Coiled current loop\tabularnewline
T. W. Marshall&
1963&
Random Electrodynamics\tabularnewline
R. K. Nesbet&
1971&
Semiclassical radiation theory\tabularnewline
J. des Cloizeaux&
1973&
Collective Fermi field excitation\tabularnewline
A. M. Arthurs&
1979&
On electromagnetic field quantization\tabularnewline
R. Woodriff \& C. Graden&
1980&
Conceptual zwitter oscillations\tabularnewline
T. H. Boyer&
1980&
Stochastic Electrodynamics\tabularnewline
A. H. Piekara\cite{Piekara}&
1982&
Electromagnetic field self-trapping\tabularnewline
M. W. Evans \emph{et al.}&
1995$\pm$&
Longitudinal magnetic field\tabularnewline
J. E. Sipe&
1995&
Photon wave functions \tabularnewline
H. Marmanis&
1998&
Navier-Stokes equations\tabularnewline
D. H. Kobe&
1999&
Relativistic second quantization\tabularnewline
I. V. Lindell&
2000&
Radiation operator\tabularnewline
P. Ghose \emph{et al.}&
2001&
Relativistic quantum mechanics\tabularnewline
B I \& V B Makshantsev&
2001&
Vector Potential\tabularnewline
V. P. Dmitriyev&
2003&
Hydrodynamic vortex\tabularnewline
Hannay \& Jeffrey \cite{Hanney}&
2005&
Synchrotron electric field\tabularnewline
P. N. Kaloyerou&
2005$\mp$&
Hidden-variable quantum mechanics\tabularnewline
I. Bialynicki-Birula&
2006$\mp$&
Quantum particle\tabularnewline
D. Dragoman&
2006&
Complex electromagnetic fields\tabularnewline
D. F. Roscoe&
2006&
Generalized Maxwell equations\tabularnewline
R. R. A. Syms \cite{Syms}&
2006&
Magneto-inductive waveguide\tabularnewline
B. J. Smith \& M. G. Raymer&
2007&
Photon wave functions\tabularnewline
\hline
\end{tabular}\par\end{centering}
\end{table}
\par\end{center}

\newpage{}

\begin{center}%
\begin{table}

\caption{Photon Characteristics\label{Table:PhotonCharacteristics}}

$^{*}$Cosmic Microwave Background at 3K equals 0.26 mev.

\begin{centering}\begin{tabular}{cccc}
\hline 
Energy&
Frequency &
Wavelength &
Intensity $\left(\frac{h\nu}{\lambda^{2}}\right)$\cite{Dirac}\tabularnewline
 $h\nu$&
$\nu$ Hz &
$\lambda$ cm &
W/(m$^{\text{2}}$$\Delta\nu$)\tabularnewline
\hline
\hline 
1Tev &
0.241E27 &
12.4E-17 &
1.0E29\tabularnewline
1Gev \cite{DeRoeck}&
0.241E24&
12.4E-14&
1.0E20 \tabularnewline
1Mev &
0.241E21&
12.4E-11&
1.0E11\tabularnewline
1Kev &
0.241E18&
12.4E-8&
1.0E2\tabularnewline
1ev &
0.241E15&
12.4E-5&
1.0E-7\tabularnewline
1mev$^{*}$&
 0.241E12&
12.4E-2&
1.0E-16\tabularnewline
1$\mu$ev &
0.241E9&
12.4E1&
1.0E-25\tabularnewline
1nev &
0.241E6 &
12.4E4&
1.0E-34\tabularnewline
1pev &
0.241E3&
12.4E7&
1.0E-43\tabularnewline
1fev&
0.241&
12.4E10&
1.0E-52\tabularnewline
1aev \cite{PhaseSkip}&
0.241E-3&
12.4E13&
1.0E-61 \tabularnewline
\hline
\end{tabular}\par\end{centering}
\end{table}
\par\end{center}

\begin{center}\newpage{}%
\begin{table}

\caption{Magnetoinductive Wave Modes\label{tab:Magnetoinductive-Wave-Modes}}

\begin{raggedright}For the first three modes, the loops lie in the
paper plane. For the last mode, the loops are normal to a common axis
in the paper plane. The magnetic field is considered normal to the
loop plane. The electric field is considered to lie in the loop gap.
Mode designations are: T, transverse; L, longitudinal; M, magnetic;
E, electric.\par\end{raggedright}

\begin{centering}\begin{tabular}{ccc}
\hline 
Mode&
Phase&
Loop Orientation\tabularnewline
\hline
\hline 
TME&
$\pm\frac{\pi}{2}$&
\ldots{} C C C C C C \ldots{}\tabularnewline
TMTLE&
$\pm\frac{\pi}{2}$&
\ldots{} J J J J J J J \ldots{}\tabularnewline
TMLE&
$\pm\frac{\pi}{2}$&
\ldots{} U U U U U U \ldots{}\tabularnewline
TELM&
$\pm\frac{\pi}{2}$&
\ldots{}| | | | | |\ldots{}\tabularnewline
\hline
\end{tabular}\par\end{centering}
\end{table}
\par\end{center}
\end{document}